# A novel scheme for measuring the relative phase difference between S and P polarization in optically denser medium

Yu Peng

*School of Physics, Beijing Institute of Technology, Beijing, 100081, P. R. China*

**Abstract**

We demonstrate applications of a novel setup which is used for measuring relative phase difference between S and P polarization at oblique incidence point in optically denser medium by analyzing the relative frequency shift of adjacent axial modes of S and P resonances of a monolithic folded Fabry-Perot cavity (MFC). The relative frequency shift of adjacent axial modes of S and P resonances of MFC is around 1.030 GHz for the confocal MFC cavity and 0.3869 GHz for the parallel MFC cavity respectively. The relative phase difference at reflection point A in optically denser medium is inferred to be around -167.4° for confocal cavity and -201.1° for parallel cavity separately. This scheme can be used to measure the relative phase difference in optically denser mediums which is out of measuring range of Ellipsometry.

## 1. Introduction

Ellipsometry is a versatile and powerful optical technique for the investigation of the complex refractive index or dielectric function tensor, which gives access to fundamental physical parameters and is related to a variety of sample properties, including morphology, glass quality, chemical composition, or electrical conductivity. The mechanism of Ellipsometry is based on exploiting the polarization transformation when a beam of polarized light is reflected from or transmitted through the interface or film and then obtains relative phase difference between S and P polarization [1]. Typically, Ellipsometry is applied only when light travels from air to an optically denser medium [2-5]. But for the case of light travelling from an optically denser medium to optically thinner medium, it doesn't work.

In this paper, we demonstrate a novel setup which is used for measuring relative phase difference between S and P at oblique incidence point in optically denser medium by analyzing the relative frequency shift of adjacent axial modes of S and P resonances of a MFC. This method can make up the measure limitation of Ellipsometry, and be used to measure relative phase difference inside the monolithic crystal where light travels from an optically denser medium to optically thinner medium.

## 2. Principle and experiments

In our experiments, the first medium is optically denser than the second, $n' = 0.687$ is relative index of refraction. $\theta_c = 86.83^\circ$ is critical value, and $\theta_B = 69^\circ$ is Brewster angle. In the case of the reflected wave, the phase changes of each component of the reflected wave will depend on the incidence magnitudes, $\theta$. To apply the Fresnel formula [6], the intensity of the light which is totally reflected is equal to the intensity of the incident light.

$$r_s = \frac{\cos\theta - i\sqrt{\sin^2\theta - n'^2}}{\cos\theta + i\sqrt{\sin^2\theta - n'^2}} = |r_s|e^{i\delta_s}$$

$$r_p = \frac{n'^2\cos\theta - i\sqrt{\sin^2\theta - n'^2}}{n'^2\cos\theta + i\sqrt{\sin^2\theta - n'^2}} = |r_p|e^{i\delta_p} \qquad (1)$$

$$\tan(\delta_s/2) = -\sqrt{\sin^2\theta - n'^2}/\cos\theta)$$

$$\tan(\delta_p/2) = -\sqrt{\sin^2\theta - n'^2}/(n'^2\cos\theta) \qquad (2)$$

According equation (1) (2), we determine the changes $\delta_s, \delta_p$, in the phases of the components of the reflected and the incident wave. The two components are seen to undergo phase jumps of different amounts. Phase difference of S and P polarization are shown in figure 1(a). Write down an expression for the relative phase difference

$$\Delta_2 = \delta_s - \delta_p = 2*\tan^{-1}(\cos\theta\sqrt{\sin^2\theta - n'^2}/\sin^2\theta) \qquad (3)$$

Between these two values there lies the maximum value of the relative phase difference. Relative phase difference between S and P polarization are shown in figure 1(b).

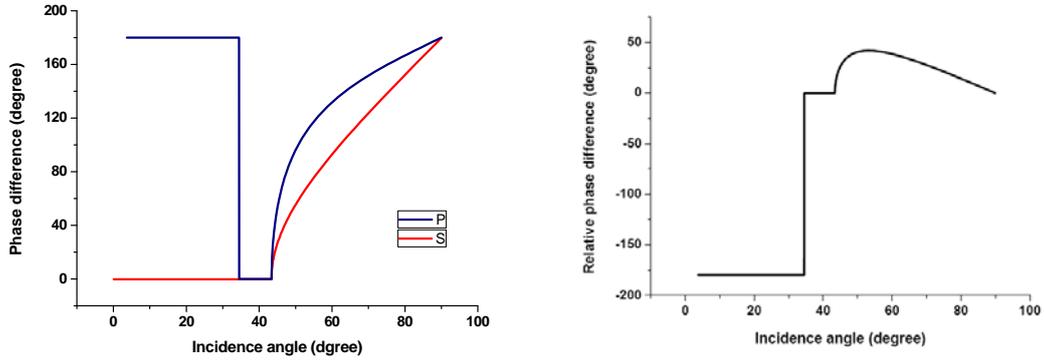

**Figure 1**. (Color online) (a) The phase difference of S polarization (red) and P polarization (blue) versus incidence angles; (b)The relative phase difference between S and P versus incidence angles.

The measure setup is shown in figure 2, for obtaining relative phase difference $\Delta$ inside optically denser medium. In this setup, a commercial MQW GaAlAs laser diode (Hitachi HL6738MG) with spectral range from 680 nm to around 695 nm operate at 689 nm and provide an output power of 5.3 mW. The threshold current is 50 mA and the laser is driven at a pump current of 60 mA. The temperature of laser diode is stabilized at room temperature by thermoelectric cooler. The laser beam is incident on the diffraction grating of 2400 grooves/mm with spectral resolution of 50 GHz. Laser frequency of laser diode with a wide spectral range is selected with diffraction grating. And resonance signal appear when laser frequency is equal to resonant frequency of the MFC by controlling PZT. S and P resonance are separately acquired by adjusting the half-wave plate, which is used for changing the polarization direction of incidence. Comparing to Ellipsometry [2-5], this method can be used to measure relative phase difference inside the monolithic where light travels from an optically denser medium to optically thinner medium.

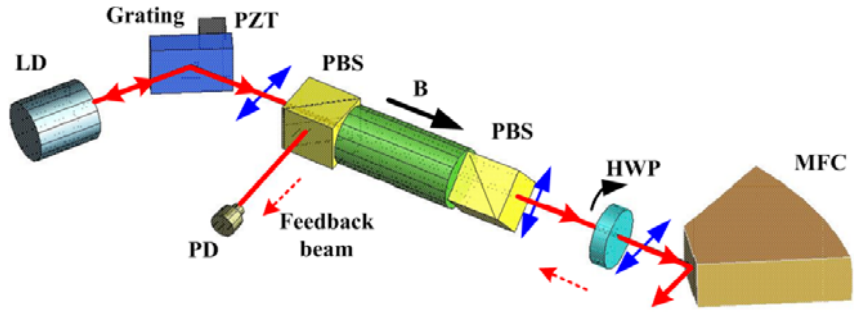

**Figure 2**. (Color online) The setup for obtaining relative phase difference inside optically denser medium of MFC: LD, laser diode; HWP, half-wave plate; PZT, piezoelectric transducer; MFC, monolithic folded F-P confocal cavity; PD, photodetector; PBS, polarizing beam splitter; blue arrow, direction of incidence's polarization; dotted red arrow, resonant feedback

The confocal MFC made of optical quartz glass is schematically shown in figure 3(a), including two optical planes, S1, S2, and an optical spherical S3, which define a confocal F-P cavity. The coupling plane S1 has a reflectivity of 0.91 (0.88) for S (P) polarization. Plane S2 is a total internal reflection surface, and S3 is a spherical mirror with a reflectivity of 0.999. The geometric length of the MFC is 30.17 mm, designed equal to the radius of curvature of S3. The finesse of the MFC is 33 (25) for S (P) polarization. The laser beam, with an external incidence angle of 45° at point A, travels along the route of ABCBADA. Resonance feedback beam is collinear with the incident light but in the opposite propagating direction, shown in figure 3(a).

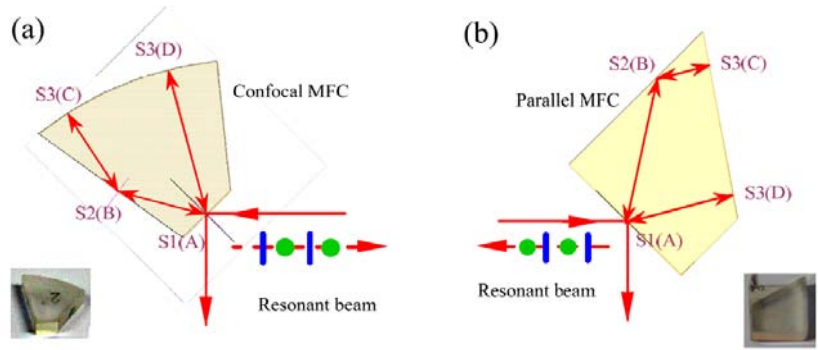

**Figure 3.** (Color online) The scheme of resonant beam of the MFC. Perpendicular(S) polarization points out of the plane of incidence (green point). Parallel (P) polarization lies parallel to the plane of incidence (blue stick). (a) confocal MFC (b) parallel MFC.

Figure 4(a) shows the changing procedure of S (green) and P (blue) component resonance of confocal MFC. In this figure, 3.413 GHz of the free spectral range is obtained by scanning laser frequency, in which higher order mode appear in the middle of the spectral range because of incidence pattern dismatching to the confocal MFC. S and P resonance are separately acquired by adjusting the half-wave plate in figure 2, which is used for changing the polarization direction of incidence. Total relative phase difference of round trip is determined by [7]

$$\Delta = 2*\pi*\Delta v*t \tag{4}$$

Where $t$ is propagation time, determined by $t = n*2*l/c$, in which n, $l$, c represent index of refraction of MFC material, geometric length of MFC cavity and speed of light in vacuum respectively. According to figure 4(a), adjacent axial modes of S and P resonances of MFC, $\Delta v_0$ is 1.030 GHz spectral shift, therefore frequency difference between S and P polarization is determined by $\Delta v = N*FSR \pm \Delta v_0$, in which N represents the number that how many FSRs the $\Delta v$ has, and FSR represents free spectrum range with the value of 3.413 GHz for the case of confocal MFC. We roughly estimate $\Delta v = N*FSR + \Delta v_0$ and $\Delta v = N*FSR - \Delta v_0$ respectively, and determine which expression is proper to ensure that expression (4) and (5) are same quantity by integer N and theoretical value of the relative phase difference at reflection point A. By equation (4), we can get the value of total relative phase difference, $\Delta$. On the other hand, total relative phase difference is expressed by [6]

$$\Delta = 2*\Delta_1 + 2*\Delta_2 + \Delta_3 + \Delta_4 \tag{5}$$

Where $\Delta_1$, $\Delta_2$, $\Delta_3$, and $\Delta_4$ are relative phase difference between S and P resonances caused by point A, B, C, and D respectively. For the total internal reflection point B, $\Delta_2$ with the value of 43.49° is calculated by the equation (3).

At normal incidence point C,D, around $\Delta_3 \approx \Delta_4 \approx -180°$ of relative phase difference are obtained. According to equation (5), therefore the relative phase difference at reflection point A, $\Delta_1$ is inferred to be around -167.4°, where contains some deviation with -180°. The deviation, we think, is caused by coating films of plane S1, shown in figure 4(b).

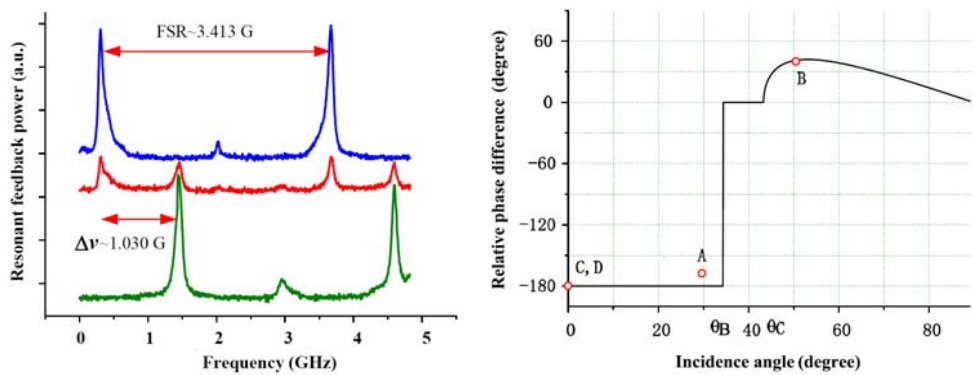

**Figure 4.** (Color online) (a) Line shape changing procedure of S (green) and P (blue) component resonance of confocal MFC, the red line is the middle status of whole procedure. (b) The relative phase difference between S and P versus incidence angles. Red mark point, A,B,C,D represent corresponding resonant reflection point inside the MFC respectively.

We apply this principle to another MFC, which is parallel cavity, shown in figure 3(b). The MFC, includes three optical planes, S1, S2, and S3, which define a parallel F-P cavity. The coupling plane S1 has a reflectivity of 0.985 (0.95) for S (P) polarization. Plane S2 is a total internal reflection surface, and S3 is a mirror with a reflectivity of 0.999. The geometric length of the MFC is 26 mm. The finesse of the MFC is 207 (61) for S (P) polarization. The right angle between S1 and S2 and the pyramidal error of S1, S2, and S3 are strictly controlled to less than 2″. The laser beam, with an external incidence angle of 46.7° at point A, travels along the route of ADABCBA. Resonance feedback beam is collinear with the incident light but in the opposite propagating direction, shown in figure 3(b). Figure 5(a) shows the changing procedure of S (green) and P (blue) component resonance of parallel MFC. In this figure, 3.97GHz of the free spectral range is obtained by scanning laser frequency, in which higher order modes appear because of incidence pattern dismatching to the parallel MFC. Adjacent axial modes of S and P resonances of MFC, $\Delta v_0$ is 0.387 GHz spectral shift. Frequency difference between S and P polarization is determined by $\Delta v = N*FSR \pm \Delta v_0$. By equation (4), we can get the value of total relative phase difference, $\Delta$. On other hand, for the total internal reflection point B, $\Delta_2 \approx 38.7°$ is calculated by the equation (3), and at normal incidence point C,D, around -180° of relative phase difference are obtained. Therefore the relative phase difference at reflection point A is inferred to be around $\Delta_1 \approx -201.1°$, shown in figure 5(b).

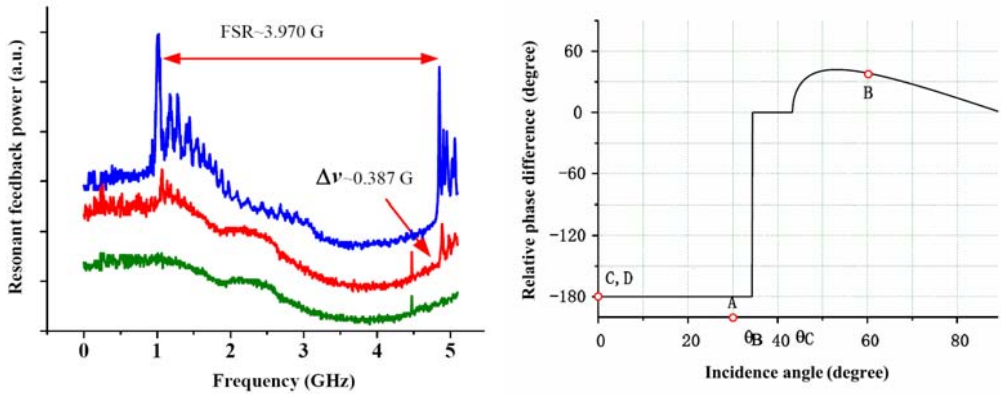

**Figure 5.** (Color online) (a) Line shape procession of S (green) and P (blue) component resonance of plane MFC, the red line is the middle status of whole procedure (b) The relative phase difference between S and P versus incidence angles. Red mark point, A,B,C,D represent corresponding resonant reflection point inside the MFC respectively.

**3. Conclusion**

In summary, the relative frequency shift of adjacent axial modes of S and P resonances of MFC is around 1.030 GHz for a case of confocal MFC cavity and 0.3869 GHz for a parallel MFC cavity respectively. The relative phase difference at reflection point A in optically denser medium is inferred to be around -167.4° for confocal cavity and -201.1° for parallel cavity separately. This setup can measure the relative phase difference in optically denser medium which exceed the measure limitation of Ellipsometry.

**Acknowledgments**

The author thanks Dr. Erjun Zang (National Institute of Metrology, China) for his useful discussions, especially his designs for MFC. This work is supported by National Major Research Program China (grants 2010CB922902).